\documentclass[onecolumn,amsmath,tightenlines,amssymb,11pt,superscriptaddress,nofootinbib]{revtex4}

\usepackage{graphicx}
\usepackage{amsmath,amssymb,amsfonts,amsthm,stmaryrd,mathtools,bm,physics,tensor}
\usepackage{color}
\usepackage{tikz}
\usepackage[normalem]{ulem}
\allowdisplaybreaks[1]

\usepackage[bookmarks,linktocpage, colorlinks=true, plainpages = false, citecolor = treegreen,  linkcolor=darkblue, urlcolor = darkblue, filecolor = blue]{hyperref} 

\def\be#1\ee{\begin{align}#1\end{align}}

\def\ba{\begin{eqnarray}}
	\def\ea{\end{eqnarray}}
\def\nn{\nonumber}

\definecolor{tealgreen}{rgb}{0.0, 0.5, 1.0}
\definecolor{darkblue}{rgb}{0., 0.4, 0.8}
\definecolor{darkgreen}{rgb}{0., 0.7, 0.3}
\definecolor{treegreen}{rgb}{0., 0.7, 0.3}

\begin{document}
	
\title{Renormalization group flows in area-metric gravity
}
\author{Johanna Borissova}
\email{jborissova@pitp.ca}
\affiliation{Perimeter Institute for Theoretical Physics, 31 Caroline Street North, Waterloo, ON, N2L 2Y5, Canada}
\affiliation{Department of Physics and Astronomy, University of Waterloo, 200 University Avenue West, Waterloo, ON, N2L 3G1, Canada}
\author{Bianca Dittrich}
\email{bdittrich@pitp.ca}
\affiliation{Perimeter Institute for Theoretical Physics, 31 Caroline Street North, Waterloo, ON, N2L 2Y5, Canada}
\author{Astrid Eichhorn}
\email{eichhorn@thphys.uni-heidelberg.de}
\affiliation{Institute for Theoretical Physics, Heidelberg University, Philosophenweg 16, 69120 Heidelberg, Germany}
\author{Marc Schiffer}
\email{marc.schiffer@ru.nl}
\affiliation{High Energy Physics Department, Institute for Mathematics, Astrophysics, and Particle Physics, Radboud University, Nijmegen, The Netherlands}

\begin{abstract}
We put forward the first analysis of renormalization group flows in an area-metric theory, motivated by spin-foam quantum gravity. Area-metric gravity contains the well-known length-metric degrees of freedom of standard gravity as well as additional shape-mismatching degrees of freedom. To be phenomenologically viable, the shape-mismatching degrees of freedom have to decouple under the renormalization group flow towards lower scales. We test this scenario by calculating the renormalization group flow of the masses and find that these are in general even more relevant than dictated by their canonical scaling dimension. This generically results in masses which are large compared to the Planck mass and thereby ensure the decoupling of shape-mismatching degrees of freedom. In addition, the latter come in a left-handed and right-handed sector. We find that parity symmetry does not emerge under the renormalization group flow. 
Finally, we extract the renormalization group flow of the Immirzi parameter from this setup and find that its beta function features zeros at vanishing as well as at infinite Immirzi parameter. 
\end{abstract}

\maketitle
\tableofcontents

\section{Introduction}\label{Introduction}

A fundamental open question in quantum gravity concerns the nature of the 
  gravitational degrees of freedom at high energies. The tentative answer to this question differs 
  in distinct quantum-gravity theories.~\footnote{This difference is a priori at the mathematical level, because it is in general not understood whether distinct quantum-gravity theories agree in their physical predictions despite being formulated in different frameworks~\cite{deBoer:2022zka}.}
  In string theory~\cite{Polchinski:1998rq,Polchinski:1998rr} the gravitational degrees of freedom arise from the vibrations of strings. In holographic approaches~\cite{Horowitz:2006ct,Hubeny:2014bla}
  gravitational degrees of freedom are encoded on the boundary of spacetime. 
  The bulk geometry is then recoverable via the scaling of the entanglement entropy of the boundary field with the area of a minimal surface cut through the bulk~\cite{Ryu:2006bv}.
  In asymptotically safe quantum gravity~\cite{Percacci:2017fkn, Reuter:2019byg} the
  gravitational degrees of freedom are carried by the metric which is assumed to describe smooth spacetimes up to arbitrarily high energies.~\footnote{There are extensions of asymptotic safety to non-metric, gravitational degrees of freedom~\cite{Daum:2010qt, Harst:2012ni, Daum:2013fu, Eichhorn:2013xr, Percacci:2013ii, Harst:2015eha}. However, these do not appear to be necessary for an asymptotically safe fixed point.} By contrast, in loop quantum gravity~\cite{Rovellibook,Thiemannbook} and its path-integral 
   formulation
  known as spin foams~\cite{Engle:2007wy,Freidel:2007py,PerezLR,HandbookSpeziale,EffSFE,EffSFL},  spacetime  
   has discrete properties.  To match to the degrees of freedom of General Relativity (GR), a transition to a field-theoretic description is necessary at some scale.
At intermediate scales between the Planckian, i.e., deep ultraviolet (UV), and  the infrared (IR) regime, the configuration space of loop quantum gravity and spin foams contains area-metric configurations instead of length-metric configurations~\cite{Dittrich:2021kzs,Dittrich:2022yoo,Borissova:2022clg,Borissova:2023yxs,Dittrich:2023ava}. Area-metric theories have also been suggested to capture phenomenological aspects of string theory~\cite{Schuller:2005yt,Schuller:2005ru,Punzi:2006hy,Punzi:2006nx,Ho:2015cza,Borissova:2024cpx}, while their impact on an asymptotically safe fixed point is not yet understood.
  
An area metric is a rank-$4$ tensor with the same symmetries as the Riemann curvature tensor~\cite{Schuller:2005yt,Schuller:2005ru,Punzi:2006hy,Punzi:2006nx,Schuller:2009hn} and therefore 20 independent components.~\footnote{A further generalization to so-called acyclic area metrics has 21 independent components~\cite{Schuller:2005yt,Schuller:2005ru,Punzi:2006hy,Punzi:2006nx,Schuller:2009hn}.}
In  a similar way as length metrics encode the lengths of tangent vectors and angles between two tangent vectors, area metrics encode the areas of tangent planes and dihedral angles between two intersecting tangent planes.
A given length metric $g$  induces an area metric $G$ defined by 
\begin{equation}
G_{\mu\nu\rho \sigma} = g_{\mu\rho}g_{\nu\sigma} - g_{\mu\sigma}g_{\nu\rho}. 
\end{equation}
However, generic area metrics cannot be expressed as induced from a length metric, because they have ten additional algebraic degrees of freedom.

A geometric understanding of these 10 additional degrees of freedom can be obtained in a discretized setting. 
 Four-dimensional triangulations are built from  four-simplices. A given four-simplex can be equipped with a flat geometry by specifying the lengths of its 10 edges.~\footnote{The geometry inside a given four-simplex is flat. Curvature is obtained by gluing four-simplices around a triangle.}  These geometric data can be mapped onto a length metric associated to the four-simplex. In loop quantum gravity, a four-simplex can be equipped with semi-classical data, given by 10 areas for the 10 triangles of the four-simplex  and two dihedral angles in each of the 5 tetrahedra which make up the four-simplex~\cite{Livine:2007vk,Freidel:2013fia}.~\footnote{More precisely, these data arise from coherent states that one associates to the boundary of four-simplices~\cite{Livine:2007vk,Freidel:2013fia}, but one can also construct a classical action for triangulations equipped with these data~\cite{Dittrich:2008va}. The resulting building blocks are essential for the construction of the spin-foam path integral~\cite{Livine:2007vk, Freidel:2007py,EffSFE}.} These 20 degrees of freedom can 
be mapped onto the 20 degrees of freedom of an area metric associated to this semiclassical simplex~\cite{Dittrich:2023ava}.

\begin{figure}[t]
	\centering
	\hspace{1.cm}
	\includegraphics[width=0.65\textwidth]{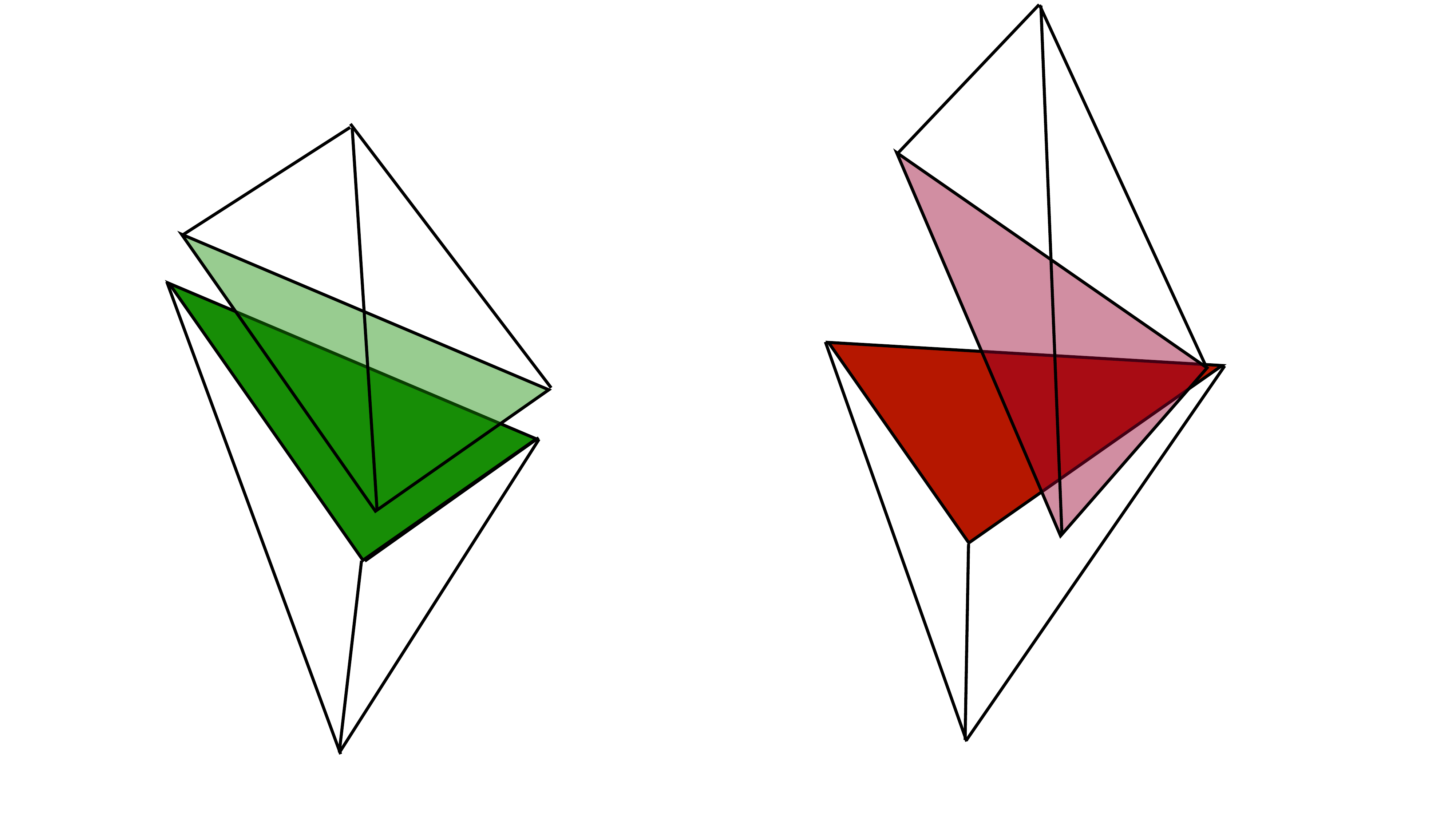}
	\caption{The basic building blocks in a four-dimensional triangulation are four-simplices. A four-simplex contains five tetrahedra. In a length geometry, pairs of tetrahedra are glued by identifying their areas and their shapes. This is shown in the configuration on the left, where the shapes of the shared green triangles match. In an area geometry, on the other hand, the shapes of the paired triangles do not need to match, while their areas match, as shown in the configuration on the right.}\label{fig:illustration}
\end{figure}

For each of the 5 tetrahedra, there are four areas for its four triangles and two dihedral angles. These specify the 6 lengths of the edges in the tetrahedron and therefore its length geometry completely. Two such tetrahedra within a given four-simplex share one triangle. The area of the triangle is uniquely specified by the geometric data of the two tetrahedra. In contrast, the two-dimensional angles in the triangle, as induced by the geometric data of the two different tetrahedra, may differ. As a result, their \emph{shapes} might not match, see~\autoref{fig:illustration}.  We thus refer to such non-length degrees of freedom in the area metric \emph shape-mismatching degrees of freedom.

This extension of the configuration space of loop quantum gravity appears due to the kinematic setup of the theory~\cite{Dittrich:2008ar, Dittrich:2010ey}.  This setup involves the simplicity constraints~\cite{Plebanski:1977zz,Capovilla:1991qb,Reisenberger:1995xh,DePietri:1998hnx,Krasnov:2010olp} which are partially second-class, i.e., anomalous~\cite{Dittrich:2008ar, Dittrich:2010ey}.~\footnote{
The notion of ``quantum anomaly" in this context refers to a non-zero commutator between the second-class part of the constraints, which is also not proportional to a constraint, despite vanishing Poisson brackets between the constraints in the continuum theory.} The second-class constraints impose shape-matching. Second-class constraints cannot be imposed sharply, as this would violate the quantum uncertainty principle. Instead, they have to be imposed weakly by allowing fluctuations which violate these constraints and  therefore lead  to area shape-mismatching. 

Area metrics appear also in the dynamics of spin foams. Using the formulation of effective spin-foams~\cite{EffSFE, EffSFL}, it has been shown that their perturbative continuum limit, evaluated on a regular lattice, leads to a (linearized) area-metric theory~\cite{Dittrich:2021kzs,Dittrich:2022yoo}. 
The  lattice theory renders the length metric degrees of freedom in the area metric massless, whereas the shape-mismatching degrees of freedom acquire Planck-scale masses. In the continuum limit linearized diffeomorphism symmetry is restored and one therefore obtains the linearized Einstein-Hilbert action  to leading order. Integrating out the massive, shape-mismatching, degrees of freedom, one obtains a higher-derivative correction, given by a Weyl-squared term~\cite{Dittrich:2022yoo}.

The mechanism of weak imposition of constraints can be mimicked also at the level of the classical action in the continuum. 
Concretely, one may construct spin-foam inspired continuum area-metric actions~\cite{Borissova:2022clg} which in turn can be understood as effective continuum actions for spin-foam dynamics. Spin foams arise from the Plebanski formulation of gravity~\cite{Plebanski:1977zz,Capovilla:1991qb,Reisenberger:1995xh,DePietri:1998hnx,Krasnov:2010olp}. Here the configuration variables are a bivector field (and a connection), instead of the more familiar tetrad field. The aforementioned simplicity constraints ensure that the bivector field is induced from a tetrad. Thus, if one  implements the simplicity constraints, one regains the Palatini formulation of GR.

By contrast, modified Plebanski theories~\cite{Krasnov:2006du, Krasnov:2008fm,Freidel:2008ku,Krasnov:2009iy, Krasnov:2009ik,Speziale:2010cf,Beke:2011mu} do not impose the simplicity constraints sharply, but suppress the corresponding degrees of freedom by a mass term, or more generally a potential. Replacing all simplicity constraints in the general modified non-chiral  Plebanski action by a potential results in more degrees of freedom than the 20 degrees of freedom of a generic area metric~\cite{Alexandrov:2008fs,Speziale:2010cf,Beke:2011mu}.  However, the simplicity constraints can be decomposed into two sets with only one imposed sharply and the remaining constraints replaced by mass terms. Such a split has been identified in~\cite{Borissova:2022clg} and results in an area-metric action. This action has been constructed explicitly in~\cite{Borissova:2022clg} to quadratic order  in a perturbative expansion in the field.

The linearized area-metric theories arising from  non-chiral Plebanski theories modified in such a way have an interesting property: they feature an additional five-dimensional shift symmetry in the kinetic term. For identical masses for all shape-mismatching degrees of freedom, the effective length-metric theory obtained after integrating out the shape-mismatching degrees of freedom is ghostfree~\cite{Borissova:2022clg, Borissova:2023yxs}. This theory corresponds to a non-local (linearized) Einstein-Weyl action, for which the spin-$2$ propagator does not exhibit additional poles beyond the massless graviton pole, unlike the general case of Stelle gravity~\cite{Stelle:1977ry}.~\footnote{The same result has been found earlier in the context of chiral modified Plebanski theories~\cite{Krasnov:2007cq,Freidel:2008ku}. In this case there are only 15 degrees of freedom and the absence of ghost poles can be explained by the five-dimensional shift symmetry of the kinetic term.}
The most general action for linearized area-metric gravity to quadratic order in fields and derivatives has been constructed in \cite{Borissova:2023yxs} (see also~\cite{Alex:2019ari} for earlier work). Generically, parity is violated in this theory, because shape-mismatching degrees of freedom occur in a left-handed and a right-handed sector, with distinct couplings. In particular, of the five free parameters of the theory, one can be identified with the parity-violating Barbero-Immirzi parameter~\cite{Borissova:2022clg}.

These very recent developments motivate a broader investigation of area-metric theories and their phenomenological viability.

 For area-metric gravity to be consistent with observations, which so far give no indication for gravitational degrees of freedom beyond those of GR, the additional shape-mismatching degrees of freedom have to decouple. The arguably simplest mechanism for decoupling makes the extra degrees of freedom too heavy to be dynamically excited in the systems that provide observational constraints. To discover whether the shape-mismatching degrees of freedom of the area metric can consistently be made heavy with masses of the order of the Planck mass, we will consider their Renormalization Group (RG) flow. Additionally, we will analyze the RG flow of parity-violating couplings in area-metric gravity. Gravitational parity violation has not been observed to date. Thus we will ask whether, given a description of gravity in terms of an area-metric in the UV, parity symmetry can be an emergent symmetry in the IR.  Finally, we will analyze the RG flow of the Immirzi parameter as a parity-violating coupling appearing for a special subclass of area-metric actions. 

 Our setup assumes a separation of scales between the fundamental (discreteness) scale in the deep UV and the Planck scale.~\footnote{The Planck scale is to be understood as the scale at which quantum gravitational degrees of freedom effectively decouple. This scale can be different from $\frac{1}{\sqrt{G_N}}= 10^{19}\, \rm GeV$, i.e., the classical value of the Planck scale, which is inferred from the low-energy dynamics of gravity. For instance, in a perturbative calculation of the beta function of the Newton coupling, the scale at which canonical scaling sets in can be pushed to significantly below $10^{19}\, \rm GeV$ by coupling a suitable number of matter fields.} In the intermediate regime between these two scales, we assume a continuum quantum-field-theoretic description of the theory with length-metric and shape-mismatching degrees of freedom. To discover the properties of this intermediate regime, we calculate RG flows driven by quantum fluctuations of the gravitational degrees of freedom.\\
In spirit, this work is related to the concept of \emph{effective asymptotic safety}~\cite{deAlwis:2019aud,Held:2020kze}, which proposes that a quantum-field theoretic description of gravity is possible beyond the Planck scale, and only breaks down at highly transplanckian scales. Within effective asymptotic safety, an RG fixed point controls the properties of the QFT at these scales and renders it approximately scale-symmetric. For the purposes of the present work, we do not assume (or systematically search for) such a fixed point, instead we assume that this regime is dominated by an interplay of length-metric and shape-mismatching degrees of freedom and the Planck scale can be understood as the scale at which shape-mismatching degrees of freedom become massive and decouple.

This article is structured as follows: In Section~\ref{Sec:PerturbativeAreaMetricGraviy}, we introduce the definition of an area metric and its algebraic decomposition into length-metric and  shape-mismatching degrees of freedom. In this section, we furthermore define our truncation ansatz for the effective average action. In Section~\ref{Sec:Propagator}, we illustrate how the area-metric propagator appearing in the flow equation is derived. Section~\ref{Sec:Constraints} spells out a set of phenomenological viability constraints that can be imposed on area-metric gravity. Section~\ref{Sec:ParitySymmMasses} analyses the scenario of decoupling of the shape-mismatching degrees of freedom in area-metric gravity. Subsequently, in Section~\ref{Sec:ParitySymmViolation} we address whether parity symmetry in the IR can emerge, starting from parity-violating initial conditions in the UV. Section~\ref{Sec:FlowBIParameter} analyses the RG flow of the Immirzi parameter as a special coupling in area-metric gravity. We finish with a discussion in Section~\ref{Sec:Discussion}.

\section{Perturbative area-metric gravity}\label{Sec:PerturbativeAreaMetricGraviy}

Unfortunately, differential geometry based on area metrics is at the present stage insufficiently developed, despite existing attempts to define connections and curvature tensors for area metrics~\cite{Schuller:2005ru,Ho:2015cza}. As mentioned above, one can also derive non-linear area-metric actions from modified Plebanski theories following~\cite{Borissova:2022clg}, but this 
has not been done yet. Thus, we extend~\cite{Borissova:2023yxs} and work perturbatively.

We consider the theory of an area metric $G$ expanded around a background configuration induced by the flat Euclidean length metric $\delta$ (used to raise and lower indices) in the form
\be\label{eq:GExpansionAroundgFlat}
G_{\mu\nu\rho\sigma} =\delta_{\mu[\rho}\delta_{\sigma]\nu} + a_{\mu\nu\rho\sigma}\,.
\ee
Here $a_{\mu\nu\rho\sigma}$ denotes the perturbations of the area metric and satisfies the same algebraic symmetries as $G$ itself, i.e.,
\be\label{eq:Symmetries}
a_{\mu\nu\rho\sigma} = - a_{\nu\mu\rho\sigma} = a_{\rho\sigma\mu\nu} \,\,\,\quad \text{and} \quad \,\,\, a_{\mu[\nu\rho\sigma]} = 0\,.
\ee
The last condition is known as the cyclicity condition. In $d=4$ dimensions, assuming the first two conditions in (\ref{eq:Symmetries}) are satisfied, it is equivalent to the condition $a_{\mu\nu\rho\sigma}\epsilon^{\mu\nu\rho\sigma}=0$, which states that this field has no totally antisymmetric component (see e.g.~\cite{Borissova:2022clg,Ho:2015cza} on the physical significance of this condition). The irreducible  $SO(4)$ representations contained in the tensor $a_{\mu\nu\rho\sigma}$ are
\be\label{eq:IrreducibleDecomposition}
a \quad \in \quad (0,0) \oplus (1,1) \oplus (2,0) \oplus (0,2) \,.
\ee
Concretely, we parametrize the area-metric perturbations analogously as in the Ricci-Weyl decomposition of the Riemann curvature tensor 
\ba\label{eq:AParametrization}
a_{\mu\nu\rho\sigma} & \equiv &  \delta_{\mu[\rho}\delta_{\sigma]\nu} h +2 \qty(\delta_{\mu[\rho}\hat{h}_{\sigma]\nu} - \delta_{\nu [\rho} \hat{h}_{\sigma]\mu}) + {\omega^+}_{\mu\nu\rho\sigma} + {\omega^-}_{\mu\nu\rho\sigma}  \\
&=& 2\qty(\delta_{\mu[\rho}h_{\sigma]\nu} - \delta_{\nu [\rho} h_{\sigma]\mu}) + {\omega^+}_{\mu\nu\rho\sigma} + {\omega^-}_{\mu\nu\rho\sigma}  \,.
\ea
Here $h$ is a scalar proportional to the trace of $a_{\mu\nu\rho\sigma}$ and $\hat{h}_{\mu\nu}$ is a symmetric and traceless $(\hat{h}_{\mu\nu}\delta^{\mu\nu} = 0)$  tensor which can be combined into the symmetric tensor 
\be\label{eq:hParametrization}
h_{\mu\nu} = \hat{h}_{\mu\nu} + \frac{1}{4}{\delta}_{\mu\nu}h\,.
\ee
The Weyl components of the area-metric perturbation ${\omega^\pm}_{\mu\nu\rho\sigma}$ are traceless (${\omega^{\pm}}_{\mu\nu\rho\sigma}\delta^{\mu\rho}=0$) and selfdual (anti-selfdual), i.e.,
\be\label{eq:Duality}
\frac{1}{2} \epsilon\indices{_\mu_\nu^\alpha^\beta} \, {\omega^\pm}_{\alpha\beta\rho\sigma} = \pm \,  {\omega^\pm}_{\mu\nu\rho\sigma}\,.
\ee
 The most general local and diffeomorphism-invariant Lagrangian at second order in area metric fluctuations and derivatives has been derived in~\cite{Borissova:2023yxs} and can be written in terms of the parametrization~\eqref{eq:AParametrization} as 
\be\label{eq:LagrangianA}
\mathcal{L}^{(2)} = \mathcal{L}^{(2)}_{\text{EH}}[h_{\mu\nu}] + \sum_{\pm}\bar{\rho}_\pm\,  \partial^\nu h^{\mu\rho}\,\partial^\sigma {\omega^\pm}_{\mu\nu\rho\sigma} + \frac{1}{2}\partial_\alpha{\omega^\pm}_{\mu\nu\rho\sigma}\, \partial^\alpha  {\omega^{\pm}}^{\mu\nu\rho\sigma}  + \frac{1}{2} \bar{m}_{\pm}^{2}\,   {\omega^\pm}_{\mu\nu\rho\sigma} \, {\omega^{\pm}}^{\mu\nu\rho\sigma} \,,
\ee
where $\mathcal{L}^{(2)}_{\text{EH}}$ denotes the linearized Einstein action at second order in the perturbations of the length metric $g_{\mu\nu}$ around flat space,
\be
g_{\mu\nu} = \delta_{\mu\nu} + h_{\mu\nu}\,.
\ee
We refer to the symmetric tensor $h_{\mu\nu}$ as the metric degrees of freedom of the area metric $a_{\mu\nu\rho\sigma}$. The area-metric Lagrangian~\eqref{eq:LagrangianA} contains mass terms for the non-metric degrees of freedom of the area metric ${\omega^\pm}_{\mu\nu\rho\sigma}$, which do not break the invariance under linearized diffeomorphisms. Parity is violated, iff $\bar{c}_+ \neq \bar{c}_-$ for any couplings $\bar{c}_\pm$. 

In the main part of this work we will analyze aspects of the RG flow for the couplings in area-metric gravity. We focus on the running of the interaction couplings $\bar{\rho}_\pm$ and the masses $\bar{m}^2_\pm$ due to gravitational interactions. To that end, as shown in the Appendix~\ref{Appendix:ThirdOrderTerms}, at third order in area-metric perturbations we have to take into account the following two types of contributions to Lagrangian~\eqref{eq:AParametrization}
\be
h\, {\omega^\pm}_{\mu\nu\rho\sigma}\,{\omega^{\pm}}^{\mu\nu\rho\sigma} \,\,\, \quad \text{and} \quad \,\,\, h^{\mu\nu}\,h^{\rho\sigma}\,{\omega^{\pm}}_{\mu\rho\nu\sigma}\,.
\ee
These provide the interaction vertices for the derivation of the beta functions. We add these two terms with a priori independent couplings $\bar{\alpha}_\pm$ and $\bar{\beta}_\pm$ to the Lagrangian~\eqref{eq:LagrangianA} and consider the following ansatz for the action as a starting point
\ba\label{eq:Gammak}
S \equiv S^{(2)}[h_{\mu\nu}] &+& \int \dd[4]{x}  \sum_{\pm}\big( \,\bar{\rho}_{\pm} \,\partial^\nu h^{\mu\rho} \,\partial^\sigma {\omega^\pm}_{\mu\nu\rho\sigma} + \frac{1}{2}\partial_\alpha{\omega^\pm}_{\mu\nu\rho\sigma} \, \partial^\alpha  \omega^{\pm\,\mu\nu\rho\sigma}  + \frac{1}{2} \bar{m}_{\pm}^{2} \, {\omega^\pm}_{\mu\nu\rho\sigma}\, \omega^{\pm\,\mu\nu\rho\sigma} \nn\\
&{}& +{\,} \bar{\alpha}_{\pm} \, h \, {\omega^\pm}_{\mu\nu\rho\sigma}\, \omega^{\pm\,\mu\nu\rho\sigma} + \bar{\beta}_{\pm} \, h^{\mu\nu}h^{\rho\sigma} {\omega^\pm}_{\mu\rho\nu\sigma} \big)\,.
\ea
We do not find a relation among $\bar{\alpha}_\pm$ and $\bar{\beta}_\pm$ by requiring diffeomorphism symmetry to cubic order in the field expansion; it is not excluded that a relation follows at higher order. 

The part of the action quadratic in the metric fluctuations $h_{\mu\nu}$ consists of the Einstein action expanded to second order in $h_{\mu\nu}$, plus a term which gauge-fixes the fluctuations with respect to the flat background $\delta_{\mu\nu}$, i.e.,
\ba
S^{(2)}[h_{\mu\nu}] &\equiv & S_{\text{EH}}^{(2)} [h_{\mu\nu}] + S_{\text{gf}}^{(2)} [h_{\mu\nu}] \nn\\
 &=& - \frac{1}{16 \pi\, G_{\mathrm{N}}} \int \dd[4]{x} \eval{\sqrt{g} R }_{\mathcal{O}(h^2)} + \frac{1}{ 32 \pi \,G_{\mathrm{N}}\,\alpha_h} \int \dd[4]{x}\qty(\partial^\rho h_{\rho \mu} - \frac{1+\beta_h}{4}\partial_\mu h)^2\,.
\ea
Here $\alpha_h$ and $\beta_h$ are gauge parameters and $G_{\mathrm{N}}$ is the dimensionful Newton coupling. We choose  $\beta_h \rightarrow \alpha_h \rightarrow 0$ in the following. Faddeev-Popov ghost terms  do not contribute to the flow of the masses and the couplings $\bar{\alpha}_\pm$, $\bar{\beta}_\pm$ and $\bar{\rho}_{\pm}$ and we thus neglect them. After expanding the Einstein action to second order in $h$, we rescale the graviton field and fields $\omega^\pm$ in Eq.~\eqref{eq:Gammak} according to
\begin{align}
\label{eq:hRescaling}
h_{\mu\nu} &\to  \sqrt{16 \pi\, G_{\mathrm{N}} \, Z_h} \, h_{\mu\nu}\,,\\
{\omega^\pm}_{\mu\nu\rho\sigma} & \to  \sqrt{Z_{\omega_{\pm}}}\, {\omega^\pm}_{\mu\nu\rho\sigma}\,,
\end{align}
where $Z_h$ and $Z_{\omega^\pm}$ denote the wave function renormalizations of $h_{\mu\nu}$ and ${\omega^\pm}_{\mu\nu\rho\sigma}$, respectively. The rescaled field $h_{\mu\nu}$ has canonical dimension one. The fields ${\omega^\pm}_{\mu\nu\rho\sigma}$ have canonical dimension one  before and after the rescaling.

\section{Area-metric propagator}\label{Sec:Propagator}
To evaluate the scale dependence of the couplings, we employ the Functional Renormalization Group (FRG)~\cite{Wetterich:1992yh, Morris:1993qb, Ellwanger:1993mw, Reuter:1996cp}. The FRG relies on the scale-dependent effective action $\Gamma_k$ which includes the quantum effects of modes with momenta $p^2>k^2$, where $k$ acts as an IR cutoff. Therefore, $\Gamma_k$ interpolates between the classical action where no quantum fluctuations are integrated out (at $k\to\infty$), and the full quantum effective action $\Gamma$, where all quantum fluctuations are integrated out (at $k\to0$). The FRG provides a differential equation for $\Gamma_k$ which describes how the system changes when integrating out modes with momenta between $k$ and $k-\delta k$. This flow equation reads
\be\label{eq:GammakFlowEquation}
k \partial_k \Gamma_k = \frac{1}{2}\text{Tr}\qty[\qty(\Gamma_k^{(2)}+ \mathcal{R}_k)^{-1} k \partial_k \mathcal{R}_k]\,,
\ee
where $\Gamma^{(2)}_k$ is the second functional derivative of $\Gamma_k$ with respect to the fields, and  $\mathcal{R}_k$ is a regulator functional which implements the shell-by-shell integration of modes, and ensures finiteness of the flows. The trace involves a trace over all continuous and discrete variables of the system. 
The beta function of couplings can be extracted from~\eqref{eq:GammakFlowEquation} via projection onto the suitable field content. For reviews and introductions to the FRG see~\cite{Berges:2000ew, Pawlowski:2005xe, Gies:2006wv, Delamotte:2007pf, Rosten:2010vm, Braun:2011pp, Reuter:2012id, Dupuis:2020fhh, Reichert:2020mja}.

In the following, we take the classical action $S$ defined in Eq.~\eqref{eq:Gammak} as an ansatz for $\Gamma_k$, i.e., we set $\Gamma_k=S$, and promote all couplings to scale-dependent couplings. 
Our choice of truncation is based on an expansion in mass dimension of couplings, together with a restriction to three-point vertices. This is partially guided by the assumption of near-perturbative RG flows, which are dominated by canonically relevant couplings; and in part guided by pragmatism for our very first study of area-metric RG flows. We refrain from introducing a larger number of couplings, which would result in (even more) unwieldy beta functions.

We choose a spectrally adjusted regulator given by
\begin{equation}
\mathcal{R}_k(p^2)=\frac{k^2}{p^2}\, r_k\bigg(\frac{p^2}{k^2}\bigg)\,\Gamma^{(2)}_k \,,
\end{equation}
where $r_k$ is the so-called shape-function, which we choose to be of Litim-type~\cite{Litim:2001up} with
\begin{equation}
r_k(x)=(1-x)\,\Theta(1-x)\,.
\end{equation}
These choices simplify the evaluation of the momentum integrals included in the trace of the flow equation, and result in analytical beta functions for all couplings.

As a first step of evaluating the flow equation~\eqref{eq:GammakFlowEquation}, we compute the regularized propagator $\qty({\Gamma_k^{(2)}} + \mathcal{R}_k)^{-1}$. Here ${\Gamma_k^{(2)}} $ is a matrix in field space spanned by $\Psi \equiv (h_{\mu\nu},{\omega^+}_{\mu\nu\rho\sigma},{\omega^-}_{\mu\nu\rho\sigma})$,
\ba\label{eq:Gamma2}
\Gamma_k^{(2)} \equiv \eval{\frac{\delta^2 \Gamma_k}{\delta \Psi_i \delta \Psi_j}}_{\Psi = 0} =
\begin{pmatrix}
	{\Gamma_k^{(2)}}_{hh} & {\Gamma_k^{(2)}}_{h\omega^+}  & {\Gamma_k^{(2)}}_{h\omega^-} \\
		\qty[{\Gamma_k^{(2)}}_{h\omega^+}]^{T} & {\Gamma_k^{(2)}}_{\omega^+\omega^+}  & 0 \\
		\qty[{\Gamma_k^{(2)}}_{h\omega^-}]^{T}  & 0  & {\Gamma_k^{(2)}}_{\omega^-\omega^-} \\
\end{pmatrix}\,.
\ea
To express the separate entries of $\Gamma_k^{(2)}$ we rewrite the action in Eq.~\eqref{eq:Gammak} using projectors $\Pi^\pm$ and $\Pi^{\text{L}}$ onto the shape-mismatching and metric components of $a_{\mu\nu\rho\sigma}$ introduced in~\cite{Borissova:2022clg,Borissova:2023yxs}, see Appendix~\ref{Appendix:Projectors}, and write
\begin{align}
a_{\mu\nu\rho\sigma} =& \,\mathbb{L}^{\lambda\tau}_{\mu\nu\rho\sigma} \, h_{\lambda\tau} + {\omega^+}_{\mu\nu\rho\sigma} + {\omega^-}_{\mu\nu\rho\sigma} \\
\equiv & \, {\Pi^{\text{L}}}_{\mu\nu\rho\sigma,\alpha\beta\gamma\delta} \, a^{\alpha\beta\gamma\delta} +{\Pi^+}_{\mu\nu\rho\sigma,\alpha\beta\gamma\delta} \, a^{\alpha\beta\gamma\delta} +{\Pi^-}_{\mu\nu\rho\sigma,\alpha\beta\gamma\delta} \, a^{\alpha\beta\gamma\delta}\,.
\end{align} 
Here $\mathbb{L}^{\lambda\tau}_{\mu\nu\rho\sigma} = 2 \delta_{\mu [\rho} \delta_{\sigma]}^{(\lambda} \delta^{\tau)}_{\nu} - 2 \delta_{\nu [\rho} \delta_{\sigma]}^{(\lambda} \delta^{\tau)}_{\mu}$ is the identity with symmetries of the Riemann tensor, which extracts the length-metric degrees of freedom from $a$. The projectors $\Pi^\pm$ act as the identity on~$\omega^\pm$. Together with the projector $\Pi^{\text{L}}$, which projects onto the length-metric part of $a$, $\Pi^\pm$ sum to the identity on the space of cyclic area metrics.

Let us now write the quadratic part of the action~\eqref{eq:Gammak} in momentum space using the above definitions.  We denote the part quadratic in $h_{\mu\nu}$ by $\mathcal{E}_{\text{PF+gf}}$. It consists of the Pauli-Fierz operator plus the gauge fixing contribution, i.e., the standard length-metric part of the two-point function. Altogether we obtain
\ba
\mathcal{L}_k^{(2)} &=& h_{\mu\nu}\, \mathcal{E}_{\text{PF+gf}}^{\mu\nu \rho\sigma}\, h_{\rho\sigma} +\sum_\pm \rho_{\pm} \, p^\nu p^\sigma h^{\mu\rho} \, {\omega^\pm}_{\mu\nu\rho\sigma} + \frac{1}{2}  \qty(p^2 + \bar{m}_\pm^2)\,{\omega^\pm}_{\mu\nu\rho\sigma} \, \omega^{\pm\,\mu\nu\rho\sigma}\\
&=& h_{\mu\nu} \, \mathcal{E}_{\text{PF+gf}}^{\mu\nu \rho\sigma} \, h_{\rho\sigma}  + \sum_{\pm} h_{\mu\nu}\, \qty(\rho_\pm \, \mathbb{I}\indices{^\mu^\nu_{\mu'}_{\nu'}}\, \delta\indices{^{\mu'}_{\alpha'}} \, \delta\indices{^{\nu'}_{\gamma'}}\,   p_{\beta'} p_{\delta'} \, {\Pi^\pm}\indices{^{\alpha'}^{\beta'}^{\gamma'}^{\delta'}_\alpha_\beta_\gamma_\delta}){\omega^\pm}^{\alpha\beta\gamma\delta} \nn\\
&{}& + \, {\omega^\pm}_{\mu\nu\rho\sigma}\qty(\frac{1}{2}\qty(p^2 + \bar{m}_\pm^2)\, {\Pi^{\pm}}\indices{^{\mu}^{\nu}^{\rho}^{\sigma}_{\alpha}_{\beta}_{\gamma}_{\delta}}){\omega^\pm}^{\alpha\beta\gamma\delta}\,.
\ea
From here we can directly read off the matrix entries
\ba 
\qty({\Gamma_k^{(2)}}_{h\omega^\pm})^{\mu\nu\,\alpha\beta\gamma\delta}  &\equiv & \eval{\frac{\delta^2 \Gamma_k}{\delta {\omega^\pm}_{\alpha\beta\gamma\delta}\, \delta h_{\mu\nu}}}_{\Psi = 0} = \rho_\pm \, \mathbb{I}^{\mu\nu\mu'\nu'}\delta\indices{_{\mu'}^{\alpha'}}\delta\indices{_{\nu'}^{\gamma'}}p^{\beta'} p^{\delta'}\, {\Pi^\pm}\indices{_{\alpha'}_{\beta'}_{\gamma'}_{\delta'}^\alpha^\beta^\gamma^\delta}  \,,\\ 
\qty({\Gamma_k^{(2)}}_{\omega^\pm\omega^\pm})^{\mu\nu\rho\sigma\,\alpha\beta\gamma\delta} & \equiv & \eval{\frac{\delta^2 \Gamma_k}{\delta {\omega^\pm}_{\alpha\beta\gamma\delta} \,\delta {\omega^\pm}_{\mu\nu\rho\sigma}}}_{\Psi = 0} = \qty(p^2 + \bar{m}_\pm^2)\Pi^{\pm\,\mu\nu\rho\sigma\,\alpha\beta\gamma\delta} \,.
\ea
The regularized propagator $G$ is defined by
\ba
\qty(\Gamma_k^{(2)}+\mathcal{R}_k) \cdot G \equiv \qty(\Gamma_k^{(2)}+\mathcal{R}_k) \cdot
\begin{pmatrix}
G_{hh} & G_{h\omega^+}  &G_{h\omega^-} \\
	\qty[G_{h\omega^+}]^{T} & G_{\omega^+\omega^+}  & G_{\omega^+\omega^-} \\
	\qty[G_{h\omega^-}]^{T}  & \qty[G_{\omega^+\omega^-} ]^T  & G_{\omega^-\omega^-}  \\
\end{pmatrix} =
\begin{pmatrix}
	\mathbb{I}_{hh} & 0&0\\
0 & \mathbb{I}_{\omega^+\omega^+}  & 0 \\
0  & 0 & \mathbb{I}_{\omega^-\omega^-}   \\
\end{pmatrix} \,.
\ea

 Using the projectors in each subsector, the entries of the propagator can be computed as
\ba
G_{hh\phantom{\alpha\beta}\rho\sigma}^{\phantom{hh}\mu\nu} &=&F_{\mathrm{TT}}\, {\Pi^{\mathrm{TT}}}_{\phantom{\,\mu\nu}\rho\sigma}^{\,\mu\nu} + F_0\, {\Pi^{0}}_{\phantom{\,\mu\nu}\rho\sigma}^{\,\mu\nu} \,,\\
G_{h\omega^{\pm}\phantom{\mu\nu}\alpha\beta\gamma\delta}^{\phantom{ h\omega_{\pm}} \mu\nu} &=&  F_{h\omega^{\pm}}\, p_{\rho'}\,p_{\sigma'}\,{\Pi^{\pm}}\indices{^{\mu}^{\rho'}^{\nu}^{\sigma'}_{\alpha}_{\beta}_{\gamma}_{\delta}}\,,\\
G_{\omega^{\pm}\omega^{\pm}\phantom{\mu\nu\rho\sigma}\alpha\beta\gamma\delta}^{\phantom{\omega_{\pm}\omega_{\pm}}\mu\nu\rho\sigma} &=& F_{\omega^{\pm}}\,{\Pi^{\pm}}\indices{^{\mu}^{\nu}^{\rho}^{\sigma}_{\alpha}_{\beta}_{\gamma}_{\delta}}\,,\\
G_{\omega_{\pm}\omega_{\mp}\phantom{\mu\nu\rho\sigma}\alpha\beta\gamma\delta}^{\phantom{w_{\pm}w_{\pm}}\mu\nu\rho\sigma} &=& F_{\omega_\mathrm{mix}}\, \frac{p_{\rho'}\,p_{\sigma'}\,p^{\mu'}\,p^{\nu'}}{(p^2)^2}\,
{\Pi^{\pm}}\indices{^{\mu}^{\nu}^{\rho}^{\sigma}_{\mu'}_{\alpha'}_{\nu'}_{\beta'}}\,
{\Pi^{\mp}}\indices{^{\rho'}^{\alpha'}^{\sigma'}^{\beta'}_{\alpha}_{\beta}_{\gamma}_{\delta}}\,,
\ea
with $\Pi_{\text{TT}}$ and $\Pi_0$ the standard projectors onto the transverse-traceless and scalar component of a symmetric rank-2 tensor, which are provided explicitly in Appendix~\ref{Appendix:Projectors}.
The coefficients $F_i$ are functions of $p^2$, the shape function $r_k$, the couplings $\rho_{\pm}$ as well as the masses $\bar{m}^2_{\pm}$, which can be explicitly computed.

The propagator contains a mixing term between the two modes $\omega^+$ and $\omega^-$. This mixing is generated by the inversion of the matrix in Eq.~\eqref{eq:Gamma2} in field space. We do not extend our truncation of $\Gamma_k$ to introduce such a mixing from the outset, because this would introduce a new coupling of negative mass dimension, which we consider of higher order in the expansion scheme underlying our truncation.
\\

We now absorb the powers of $G_{\mathrm{N}}$ arising from the rescaling in Eq.~\eqref{eq:hRescaling} and define
\begin{equation}
\hat{\rho}_{\pm}= \bar{\rho}_{\pm}\,\sqrt{16\pi \, G_{\mathrm{N}}}\,,\quad \hat{\alpha}_{\pm}=\bar{\alpha}_{\pm}\,\sqrt{16\pi \, G_{\mathrm{N}}}\,,\quad \hat{\beta}_{\pm}=16\pi \, G_{\mathrm{N}}\,\bar{\beta}_{\pm}\,.
\end{equation}
 This results in a standard form of the kinetic mixing term ${\Gamma_k^{(2)}}_{h\omega^{\mp}}$.
Since beta functions are most conveniently computed for dimensionless versions of the couplings, we introduce the dimensionless couplings
\ba
g = k^2 \, G_{\mathrm{N}} \,,\quad \rho_{\pm} =  \hat{{\rho}}_{\pm} \,,\quad m_{\pm}^2 = k^{-2} \, \bar{m}_\pm^2\,,\quad \alpha_\pm= k^{-1} \, \hat{\alpha}_{\pm}\,, \quad \beta_\pm= k^{-1} \, \hat{\beta}_{\pm}\,.
\ea

\begin{table}
\begin{tabular}{c|c|c|c}
n-point function & couplings in original action & parity-symmetric subspace & parity-violating subspace\\ \hline \hline
$h \omega \omega$ vertex  & $\alpha_{\pm}$ & $\sigma_{\alpha} = 
\alpha_+ + \alpha_-$ & $\delta_{\alpha} = \alpha_+ - \alpha_-$ \\ \hline
$h h \omega$ vertex & $\beta_{\pm}$ & $\sigma_{\beta} = \beta_+ + \beta_-$ & $\delta_{\beta} = \beta_+ - \beta_-$ \\ \hline
$\omega$ mass term & $m_{\pm}^2$ & $\sigma_{m^2}= m_+^2 + m_-^2$ &   $\delta_{m^2}= m_+^2 - m_-^2$ \\ \hline
$h\omega$ mixing & $\rho_{\pm}$ & $\sigma_{\rho} = \rho_+ + \rho_-$ & $\delta_{\rho} = \rho_+ - \rho_-$\\ \hline
\end{tabular}
\caption{\label{Tab:couplings} We list the couplings we analyze in this work. $\sigma$'s and $\delta$'s are used explicitly in Sec.~\ref{Sec:ParitySymmViolation}.}
\end{table}

\section{Phenomenological viability constraints on the couplings and RG flows towards the IR}\label{Sec:Constraints}

Area-metric gravity has more degrees of freedom than GR. To be consistent with observations, which so far give no indication for gravitational degrees of freedom beyond those of GR, these extra degrees of freedom have to decouple. The arguably simplest mechanism for decoupling makes the shape-mismatching degrees of freedom too heavy to be dynamically excited at experimentally accessible energies. To discover whether the  shape-mismatching degrees of freedom $\omega^{\pm}$ can consistently be made heavy, with masses of the order of the Planck mass, we consider their RG flow. If the dimensionless mass squared $m^2$ scales with a scaling dimension $\Delta_{m^2}$, i.e., $m^2 \sim k^{\Delta_{m^2}}$, then the dimensionful mass squared $\bar{m}^2$ scales as $\bar{m}^2 \sim k^2 \cdot k^{\Delta_{m^2}}$. Canonical scaling is $\Delta_{m^2}=-2$, such that the dimensionful mass is constant. 

In the absence of interactions masses follow this canonical scaling. However, in the presence of gravitational fluctuations masses acquire an anomalous scaling dimension $\gamma_{m^2}$ such that $\Delta_{m^2} = -2 + \gamma_{m^2}\neq -2$. The dimensionful mass in the presence of an anomalous scaling dimension scales as
\begin{equation}
\bar{m}^2(k) \sim k^2 \cdot m^2(\Lambda_{\rm UV
}) \left(\frac{k}{\Lambda_{\rm UV}}\right)^{\Delta_{m^2}} \sim m^2(\Lambda_{\rm UV}) \cdot \Lambda_{\rm UV}^2\cdot \left(\frac{k}{\Lambda_{\rm UV}}\right)^{\gamma_{m^2}}.
\end{equation}
Herein, $\Lambda_{\rm UV}$ is a UV-cutoff scale at which the initial condition for the RG flow, $m^2(\Lambda_{\rm UV})$, is fixed. In the present setting this is the scale at which a more fundamental quantum-gravity description gives rise to an effective quantum field theory of the type we consider here. 

 Under the impact of length-metric fluctuations, the anomalous scaling dimension for the mass is generically positive for spin-1/2 and spin-0 fields, see~\cite{Eichhorn:2016vvy} and~\cite{Narain:2009fy,Wetterich:2016uxm,Eichhorn:2017als}, respectively. 
This means that the mass is less relevant than expected canonically. Therefore, the dimensionful mass at some IR scale $k_{\rm IR}$ is smaller than the UV cutoff scale by the factor $\left(\frac{k_{\rm IR}}{\Lambda_{\rm UV}}\right)^{\gamma_{m^2}}$. Generically, it should be assumed that the anomalous scaling holds over a finite range of scales and $k_{\rm IR}$ cannot be taken to zero in this expression.

For area-metric degrees of freedom the analogous question has not yet been investigated. It is, however, crucial for the phenomenological viability of the theory that shape-mismatching degrees of freedom decouple. The simplest way for them to do so is to become very massive.

Thus, in the next Section~\ref{Sec:ParitySymmMasses}
we explore where in the parameter space spanned by the couplings in our truncation the masses of the  shape-mismatching degrees of freedom become RG irrelevant, such that $\Delta_{m^2}>0$.
In practice, given a beta function for the square of the mass, we subtract the canonical term and determine whether the remaining term is positive, such that the mass tends to decrease towards the IR, or negative, such that the mass tends to increase towards the IR. The latter is the ``phenomenologically safe" option.

 In addition, we test whether decoupling is also possible through a suppression of the interactions between $h$ and $\omega^{\pm}$.

Subsequently, in Section~\ref{Sec:ParitySymmViolation} we analyze parity symmetry. This is phenomenologically motivated by the absence of observational indications for gravitational violations of parity~\cite{Zhu:2023rrx,Yunes:2025xwp}. If the $\omega^+$-sector has different couplings from the $\omega^-$-sector, the effective theory for $h$, obtained after integrating out $\omega^{\pm}$, is likely to exhibit parity violation. Thus, we aim at investigating whether the symmetry $\omega^{\pm} \rightarrow \omega^{\mp}$ is an emergent symmetry, i.e., whether deviations from this symmetry are suppressed by the RG flow.

\section{Do non-metric degrees of freedom decouple?}\label{Sec:ParitySymmMasses}

According to our discussion in Section~\ref{Sec:Constraints}, the question of primary interest is whether there exist regions in the space of couplings where the masses $m_\pm^2$ of the non-metric degrees of freedom become more relevant through quantum fluctuations. If this is the case, we can state that, irrespective of the initial conditions $m_\pm^2(\Lambda_{\rm UV})$, the non-length-metric degrees of freedom decouple. On the other hand, if the masses are not rendered more relevant through quantum fluctuations, the resulting physical masses can still be large if the initial conditions are chosen appropriately. Thus, we are interested in discovering whether there is a region in coupling space, where the masses are \emph{generically} large, and not just in settings with ``natural" values of couplings, where $m_\pm^2(\Lambda)\sim \mathcal{O}(1)$ is assumed. This is motivated by the fact that the masses are currently not computable from spin-foam models, such that a generic statement that does not depend on the initial condition is of most interest. 

In an expansion of $\beta_{m_\pm^2}$ for small $m_\pm^2$ and small $\rho_\pm$, $\alpha_\pm $, $\beta_\pm$, the leading-order term is of the order $m_\pm^0$, such that $m_\pm^2$ is automatically generated by quantum fluctuations. The leading-order expansion is 
\be
\beta_{m_\pm^2} = - \frac{2 \alpha_\pm^2}{3 \pi^2} + \frac{7 \beta_\pm^2}{192 \pi^2} +\dots\,.
\ee
If we assume that the values of the couplings $\alpha_\pm$ and $\beta_\pm$ are roughly the same, then the first term in $\beta_{m_\pm^2}$ is dominant. This term is negative for either sign of $\alpha_\pm$, resulting in a growth of $m_\pm^2$. Thus, even if the initial condition for the masses is $m_\pm^2(\Lambda_{\rm UV})=0$, masses are generated by the RG flow.\\

In a next step, we investigate the behavior of $m_\pm^2$ at large $m_\pm^2$ to test whether the masses, once they are sizable, are protected from decreasing under the RG flow.

Expanding the beta functions $\beta_{m_\pm^2}$ for large $m_\pm^2$ leads to 
\ba\label{eq:Betam}
\beta_{m_\pm^2}& = &  \qty(-2  - \frac{3\beta_\pm^2}{128 \pi^2})m_\pm^2 
+\mathcal{O}\qty(m_\pm^0)
\ea
The anomalous contribution to the scaling dimension is $\gamma_{m_\pm^2} = - 3 \beta_\pm^2/(128\pi^2)$, which results in the dimensionful masses growing towards the IR, thereby  decoupling the shape-mismatching degrees of freedom.\\

This decoupling may be circumvented, if the vertex couplings between $\omega$ and $h$ become large, too.
Expanding the beta functions for the vertex couplings $\alpha_\pm$ and $\beta_\pm$
for large $m_\pm^2$ leads to
\ba
\beta_{\alpha_\pm}&=&   \qty(-1- \frac{3\beta_\pm^2}{128\pi^2})\alpha_\pm+\frac{7 \sqrt{\pi g}  \beta_\pm^2}{96\pi^{2}} +\mathcal{O}\qty(m_\pm^{-2})\,,\label{eq:BetaAlpha}\\
\beta_{\beta_\pm}&=& \qty(-1- \frac{3\beta_\pm^2}{256\pi^2})\beta_\pm +\mathcal{O}\qty(m_\pm^{-2})\label{eq:BetaBeta}\,.
\ea

Canonical scaling implies $\hat{\alpha}_\pm\sim \Lambda_{\rm UV} \sim  \hat{\beta}_\pm$, because both couplings have mass dimension one. To obtain a suppression relative to this canonical expectation, the anomalous scaling dimension has to be of opposite sign to the canonical scaling.

We find that this is not achievable for both $\alpha_\pm$ and $\beta_\pm$.
Specifically, from Eq.~\eqref{eq:BetaBeta} we see that this condition cannot be met for the coupling $\beta_\pm$ as
\be
\text{sign}\qty(-\frac{3\beta_\pm^3}{256 \pi^2}) = - \text{sign}(\beta_\pm)\,.
\ee
Thus we cannot achieve that $\beta_\pm$ is shifted towards irrelevance in the IR and within the large-$m_\pm^2$ regime. 
However, from Eq.~\eqref{eq:BetaAlpha} we see that the non-canonical part in the beta function for $\alpha_\pm$ 
satisfies
\be
\text{sign}\qty(-\frac{3 \beta_\pm^2}{128 \pi^2}\alpha_\pm + \frac{7 \sqrt{\pi g}\beta_\pm^2}{96 \pi^2}) = \text{sign}\qty(28 \sqrt{\pi g} - 9 \alpha_\pm)
\ee
If $\alpha_\pm$ is negative, this sign is positive, as is the sign of the canonical term. Thus, $\alpha_\pm$ is even more relevant than its canonical scaling implies. 
If, however,
$\alpha_\pm $ is positive, then we can achieve 
anomalous scaling opposite to canonical scaling,
provided  $\alpha_\pm < \frac{28 \sqrt{\pi g}}{9}$.  

In summary, at least one of the two couplings remains relevant and thus of the order of $\Lambda_{\rm UV}$, at least in the absence of fine-tuning of the initial condition for the coupling at $\Lambda_{\rm UV}$.
Overall, there may therefore remain an effect of the shape-mismatching degrees of freedom in the low-energy theory. One the one hand, these fields cannot propagate if their masses are large, as we find good indications for. However, on the other hand, because their coupling to $h$ can also stay large (in units of $\Lambda_{\rm UV}$), we expect that they may generate large Wilson-coefficients in the effective field theory for the length-metric degrees of freedom, presumably resulting in large higher-order curvature terms in the effective action for $h$.

An alternative decoupling mechanism consists in the couplings between shape-mismatching degrees of freedom and length-metric degrees of freedom being driven to zero even at small masses. Thus, we consider the vertex couplings $\alpha_\pm$ and $\beta_\pm$. In an expansion of $\beta_{\alpha_\pm}$ and $\beta_{\beta_\pm}$ 
$m_\pm^2$ and $\rho_\pm,\alpha_\pm,\beta_\pm$, the leading term is of the order $m_\pm^0$, i.e., a  flow for these couplings is induced even in the absence of masses for the non-metric degrees of freedom. Thus, decoupling of shape-mismatching degrees of freedom by vanishing couplings is not possible at small values of the masses.

\section{Is parity an emergent symmetry?
}\label{Sec:ParitySymmViolation}

 Under parity transformations, the selfdual and anti-selfdual fields are exchanged, i.e.,
\ba
h_{\mu\nu} \to h_{\mu\nu} \,\,\,\quad \text{and} \,\,\,\quad {\omega^{\pm}}_{\mu\nu\rho\sigma} \to {\omega^{\mp}}_{\mu\nu\rho\sigma} \,.
\ea
This is a global symmetry of the action if the couplings satisfy the following relations
\be
c_+ = c_- 
\quad \quad  
\forall \,\,\text{couplings} \,\,c_{\pm } \in \Gamma_k \,.
\ee
In addition, the RG flow satisfies this symmetry and thus $\beta_{c_+} = \beta_{c_-}$ if $c_+=c_-$.~\footnote{This contradicts the conjecture that quantum gravity breaks all global symmetries~\cite{Banks:1988yz, Banks:2010zn, Harlow:2018tng, Daus:2020vtf}. Thus, either the conjecture does not hold in all quantum-gravity settings, as emphasized e.g.~in~\cite{Borissova:2024hkc}; or our Euclidean FRG calculation does not adequately account for virtual black-hole configurations at the heart of the conjecture~\cite{Eichhorn:2024rkc}.}
 We now explore whether parity symmetry is an emergent symmetry, such that it is generated by the RG flow, even if it is violated in the deep UV. Therefore, we study the RG flow outside the symmetric subspace of the space of couplings.
To that end we introduce
 differences and sums of couplings, 
\be\label{eq:CouplingsDiffSum}
\delta_c \equiv  c_+ - c_- \quad \text{and}\quad  \sigma_{c} \equiv  c_+ + c_- \quad \quad \quad  \forall \,\,\text{couplings} \,\,c_{\pm } \in \Gamma_k \,.
\ee
The couplings $\delta_c $ parametrize the strength of parity violation, whereas $\sigma_c$ parametrize the parity-symmetric subspace. 
We expand the beta function of the mass difference $\delta_{m^2}$ to leading order in $\frac{1}{\sigma_{m^2}}$, i.e., for large sum of the masses, and to leading order in parity-violating differences. Thus, we test whether the phenomenologically interesting regime of large mass has emergent parity symmetry, which would imply $\delta_c \rightarrow 0$ for all couplings.  We obtain
\ba\label{eq:BetaDeltam}
\beta_{\delta_{m^2}} &=&  \qty(-2 - \frac{3 \sigma_{\beta}^2}{512 \pi^2} )\delta_{m^2}
- \frac{\qty(2 \sigma_\rho^2 + 9 \sigma_{m^2} - 28)\sigma_\beta}{768 \pi^2}\delta_\beta + 
\dots\,,\label{eq:BetaDeltaM}\\
\beta_{\delta_\rho} &=& \frac{3 \sigma_\beta^2}{1024 \pi^2}\delta_\rho - \qty(\frac{31 \sqrt{\pi g}}{54 \pi^2} + \frac{3 \sigma_\rho \sigma_\beta }{512\pi^2}) \delta_\beta + \dots\,,
\label{eq:BetaDeltaRho}\\
\beta_{\delta_\alpha} &=& \qty(-1 -\frac{3 \sigma_\beta^2}{512 \pi^2})\delta_\alpha - \frac{\qty(56 \sqrt{\pi g}- 9 \sigma_\alpha)\sigma_\beta}{768 \pi^2} \delta_\beta + \dots
\,,\label{eq:BetaDeltaAlpha}\\
\beta_{\delta_\beta} & = &  \qty(- 1 - \frac{9 \sigma_\beta^2}{1024 \pi^2})\delta_\beta + \dots
\label{eq:BetaDeltaBeta}\,.
\ea
The stability matrix around the Gaussian fixed point $\delta_{m^2,\, \ast}=0=\delta_{\rho,\, \ast}= \delta_{\alpha,\,\ast}= \delta_{\beta,\, \ast}$ is upper-triangular such that the critical exponents  are given by minus the linear coefficient in each beta function. These are all negative, except for $\delta_{\rho}$, such that it is the only coupling in which parity violation is dynamically kept at zero, because the leading-order term in $\beta_{\delta_{\rho}}$ is positive. Thus, parity is not an emergent symmetry.

The additional terms in each of the beta functions, which are all proportional to $\delta_{\beta}$ can have various signs. Thus, if $\delta_{\beta}$ is already non-zero, this coupling can in turn limit the further growth of some of the other couplings, at least in parts of the parameter space. Overall, however, there is no way for parity violations, once they are present, to be dynamically driven back to zero.\\
To avoid parity violation, the initial conditions at $\Lambda_{\rm UV}$ thus have to be chosen to respect parity symmetry exactly. If the parity violation considered here has undesirable phenomenological consequences, as we conjecture, then this choice of initial conditions may be the only viable one in the theory.

\section{Flow of the Immirzi parameter}\label{Sec:FlowBIParameter}

The Barbero-Immirzi parameter, or Immirzi parameter, $\gamma$~\cite{BarberoG:1994eia,Immirzi:1996di} originally appeared in the construction of the canonical loop quantum gravity variables, in particular the so-called Ashtekar-Barbero connection~\cite{AshtekarVar,BarberoG:1994eia}. It was then found that these variables arise from a canonical analysis of the Palatini (or tetrad) action~\cite{PerezCanon}, if one adds to this action the so-called Holst term~\cite{Holst:1995pc}, multiplied with the (inverse of the) Immirzi coupling. This Holst term is parity-violating but quasi-topological, that is, it does not affect the classical equations of motion.~\footnote{Up to a boundary contribution the Holst term equates to a torsion squared term.} But it induces a canonical transformation which changes the momenta from being given by the extrinsic curvature to the Ashtekar-Barbero connection. The Immirzi parameter plays therefore a central role in loop quantum gravity.~\footnote{The Immirzi parameter appears in the  discrete spectrum of spatial area operators~\cite{Rovelli:1994ge,Ashtekar:1996eg}. This has been perplexing as in the classical theory, the canonical variables for different Immirzi parameter are related by a canonical transformation, and one would thus expect theories with different Immirzi parameters to be unitarily equivalent. But this appearance of the Immirzi parameter in the spectra of operators appears less surprising when one takes into account the enlargement of the loop quantum gravity configuration space from length to area metrics discussed in Sec.~\ref{Introduction}.}

The renormalization flow of the Immirzi parameter has been studied  using functional renormalization methods applied to first order tetrad gravity~\cite{Daum:2010qt,Benedetti:2011nd,Daum:2013fu,Harst:2012ni} and second order tetrad gravity with the Holst term~\cite{Harst:2012ni}. 

Here we will investigate the flow of the Immirzi parameter within the area-metric framework. Interestingly, the Immirzi parameter plays also a crucial role~\cite{Dittrich:2012rj} in the enlargement of the loop quantum gravity configuration space from length to area metrics, which we discussed in the Introduction (Sec.~\ref{Introduction}): it appears as anomaly parameter in the commutator algebra of the simplicity constraints~\cite{Dittrich:2008ar,Dittrich:2010ey,Dittrich:2012rj}. It therefore controls how strongly the anomalous part of the simplicity constraints can be imposed, and thus how strongly the shape-mismatching degrees of freedom can be suppressed~\cite{EffSFE,EffSFL}: the smaller the Immirzi parameter, the more one can suppress the shape-mismatching degrees of freedom.  In the context employed in this paper one would suppress the shape-mismatching degrees of freedom by increasing their mass. We note however that we will treat the masses as independent parameters, and therefore do not capture this connection between the value of the Immirzi parameter and the masses of the shape-mismatching degrees of freedom.

The Holst term in the Palatini framework is characterized by being parity violating, for the quadratic area metric action we therefore identify the Holst term with the parity-violating terms~\footnote{Alternatively, see~\cite{Borissova:2022clg} for a derivation of area-metric gravity from a modified non-chiral Plebanski action (with Holst term), and~\cite{Beke:2011mu} for earlier analysis of non-chiral modified Plebanski actions.}, that is 
\be\label{eq:L2}
	\mathcal{L}^{(2)} =  \mathcal{L}^{(2)}_{\text{EH}}(h) + \frac{1}{2}\sum_\pm \gamma_\pm \, h \,\omega^\pm\, p\, p + \frac{1}{8 } \gamma_\pm \,\omega^\pm\,(p^2 + m_\pm^2)\,\omega^\pm\,,
\ee
where we have omitted tensor indices for clarity. The tensorial structures can be read off from~\eqref{eq:LagrangianA} or~\eqref{eq:Gammak}. 
The dimensionless Immirzi parameter $\gamma$ enters in~\eqref{eq:L2} explicitly through the couplings
\be\label{eq:GammaPM}
\gamma_\pm = \frac{1}{8 \pi \,G_\text{N}}\qty(1 \pm \frac{1}{\gamma})\,.
\ee
For $\gamma^{-1} \to 0$, corresponding to a suppression of the Holst term, parity violation in $\mathcal{L}^{(2)}$ is absent, if $m^2_+=m^2_-$ holds. %
For $\gamma^{-1} \to\pm \infty$, parity violation is maximal.

It is important to note, that in area-metric gravity the Holst term is not quasi-topological anymore; it does affect the equations of motion~\cite{Borissova:2023yxs}.

The definition (\ref{eq:GammaPM}) of $\gamma$ requires that  $\gamma_+ + \gamma_- = 1/(4 \pi G_\text{N})$ holds. With this requirement the action (\ref{eq:L2}) defines a special case of the general quadratic area metric actions introduced in (\ref{eq:Gammak}): the kinetic term of~\eqref{eq:L2} exhibits a $5$-parameter shift symmetry (which is broken by the mass terms), see~\cite{Borissova:2022clg,Borissova:2023yxs}.  Notably, when the mass parameters $m_\pm^2 $ are identical, the effective action for $h$ obtained after integrating out $\omega^\pm$ from Eq.~\eqref{eq:L2} is ghostfree~\cite{Borissova:2022clg,Borissova:2023yxs}.

In order to employ the same conventions as used for~\eqref{eq:Gammak}, we rescale the fields $h$ and $\omega^\pm$ as follows
\ba
h &\to & \sqrt{16 \pi \,G_\text{N} Z_h} \,h\,,\\
\omega^\pm &\to & \sqrt{ 8 Z_{\omega^\pm}/\gamma_\pm}\, \omega^\pm \label{eq:Rescalingw}\,.
\ea

Again, we redefine the couplings in $\Gamma_k$ so as to absorb all but the wave-function renormalizations appearing in front of a given term. As a consequence, the dimensionless couplings $\rho_\pm \equiv \frac{1}{2} \sqrt{16 \pi G_\text{N} \cdot 8\gamma_\pm}$ appearing in front of the $h\omega^\pm$ interaction terms satisfy
\ba
\rho_+^2 + \rho_-^2 &=& 8\,,\label{eq:RhoSqSum}\\
\rho_+^2 - \rho_-^2 &=&  \frac{8}{\gamma}\label{eq:RhoSqDifference}\,.
\ea

These two relations determine how $\gamma$ appears in the classical action. However, under the RG flow the condition~\eqref{eq:RhoSqSum} is generically not preserved. To take this into account, we introduce a coupling proportional to the sum of $\rho_+^2$ and $\rho_-^2$ and write
\ba
\rho_+^2 + \rho_-^2 &=& 8\, \sigma_{\rho^2}\,,\label{eq:RhoSqSumNew}\\
\rho_+^2 - \rho_-^2 &=&  \frac{8}{\gamma}\label{eq:RhoSqDifferenceNew}\,.
\ea

Solving Eq.~\eqref{eq:RhoSqSumNew} and Eq.~\eqref{eq:RhoSqDifferenceNew} for $\rho_+$ and $\rho_-$ in terms of $\sigma_{\rho^2}$ and $\gamma$ results in~\footnote{There are actually four sets of solutions with all four possible combinations of signs. The two mixed combinations can be excluded, because $\gamma^{-1} \rightarrow 0$ should restore parity, requiring $\rho_+=\rho_-$. Of the two remaining combinations, we can choose the positive signs without loss of generality.}
\be\label{eq:RhoPM}
\rho_+  = 
2 \sqrt{\sigma_{\rho^2} + \gamma^{-1}} \,\,\, \quad \text{and} \,\,\, \quad \rho_-  =  
2 \sqrt{\sigma_{\rho^2}  - \gamma^{-1}}.
\ee
From Eq.~\eqref{eq:RhoSqDifferenceNew}, we obtain
\be\label{eq:BetaGammaInv}
\beta_{\gamma^{-1}} = \frac{1}{4}\qty( \rho_+(\gamma^{-1})\cdot \beta_{\rho_+} - \rho_- (\gamma^{-1})\cdot \beta_{\rho_-} )\,\,\,\quad \text{where} \quad  \rho_\pm\qty(\gamma^{-1}) \equiv 2 \sqrt{\sigma_{\rho^2} \pm \gamma^{-1}}\,.
\ee
Expanding $\beta_{\rho_\pm}$ for large masses we find for $\beta_{\gamma^{-1}}$

\ba\label{eq:BetaGammaInvExpansion}
\beta_{\gamma^{-1}} &=& - \frac{ 81 \qty(\beta_+^2 \qty(\sigma_{\rho^2} + \gamma^{-1}) - \beta_-^2 \qty(\sigma_{\rho^2} - \gamma^{-1})) + 1984 \sqrt{g \pi} \qty(\beta_+ \sqrt{\sigma_{\rho^2} + \gamma^{-1}} - \beta_- \sqrt{\sigma_{\rho^2} - \gamma^{-1}})}{6912 \pi^2} \nn\\
&+& \mathcal{O}\qty(m_\pm^{-2}) \,.
\ea
 Notably the leading term in this expansion is independent of the couplings $\alpha_\pm$.
 \\
 
 Additionally, we can expand the beta function for $\sigma_{\rho^2} $ for large $m^2$, 
 \ba\label{eq:BetaRhoSqSumExpansion}
 \beta_{\sigma_{\rho^2}} &=& - \frac{ 81 \qty(\beta_+^2 \qty(\sigma_{\rho^2} + \gamma^{-1}) + \beta_-^2 \qty(\sigma_{\rho^2} - \gamma^{-1})) + 1984 \sqrt{g \pi} \qty(\beta_+ \sqrt{\sigma_{\rho^2} + \gamma^{-1}} + \beta_- \sqrt{\sigma_{\rho^2} - \gamma^{-1}})}{6912 \pi^2} \nn\\
 &+& \mathcal{O}\qty(m_\pm^{-2}) \,.
 \ea
From Eq.~\eqref{eq:BetaGammaInvExpansion} we observe that the RG flow for $\gamma^{-1}$ vanishes if we take $\beta_{\pm}\rightarrow 0$ and $m_\pm^2 \rightarrow \infty$, i.e., if we decouple the shape-mismatching degrees of freedom. Purely length-metric fluctuations do not result in a flow for $\gamma^{-1}$. 
 
 Moreover, the flow of $\gamma^{-1}$ vanishes for $\gamma^{-1}\to 0$ and $\sigma_{\rho^2} \to 0$, i.e., if the interaction couplings $\rho_\pm$ in front of the $h\omega^\pm$ terms are switched off. At this point, the derivative of the beta function $\beta_{\gamma^{-1}}$ diverges such that no well-defined critical exponent can be assigned; however, a sign can be assigned, because
 \begin{equation}
\partial_{\gamma^{-1}}\beta_{\gamma^{-1}} \Big|_{\gamma^{-1} \rightarrow 0} = -\frac{81}{6912 \pi^2} \left(\beta_+^2 - \beta_-^2 \right) - \frac{1984}{6912 \pi^2}\sqrt{g \pi}\left(\frac{\beta_+}{2 \sqrt{\sigma_{\rho^2}}}- \frac{\beta_-}{2\sqrt{\sigma_{\rho^2}}} \right).
 \end{equation}
 In the limit $\sigma_{\rho^2} \rightarrow 0$, the second term diverges and thus dominates the flow. The sign is determined by the combination $\beta_+ - \beta_-$, such that the RG flow is driven towards (away from) $\gamma^{-1}=0$ for negative $\beta_{+}+\beta_-<0$  ($\beta_{+}+\beta_->0$). 

There are also other zeros of the two beta functions (in the large-mass limit), notably the two configurations $\qty{\sigma_{\rho^2} = + \gamma^{-1}, \beta_+ = 0}$ and $\qty{\sigma_{\rho^2} = - \gamma^{-1}, \beta_- = 0}$.
For example, it is possible to realize the condition for shift symmetry in the kinetic term of the classical action, for  $(\sigma_{\rho^2} ,\gamma^{-1},\beta_+)=(1,1,0)$ or $(\sigma_{\rho^2} ,\gamma^{-1},\beta_-)=(1,-1,0)$. These two configurations are special, as in this case one of the interaction couplings $\rho_\pm$ between the length-metric sector and either the selfdual or the anti-selfdual component $\omega^\pm$ at quadratic order is switched off entirely. However, this component is still propagating, as, after applying the rescaling~\eqref{eq:Rescalingw} in~\eqref{eq:L2}, the term quadratic in $\omega^\pm$ consists of a standard kinetic plus mass term which is independent of $\gamma_\pm$.

We are also interested in the point $\gamma=0$. To investigate this point we derive $\beta_\gamma$ from $\beta_{\gamma^{-1}}$ through the relation
\begin{eqnarray}
\beta_{\gamma} &=& -\gamma^2 \beta_{\gamma^{-1}}\nonumber \\
&=&\frac{3}{256 \pi^2}  \qty(\beta_+^2 + \beta_-^2)\gamma +\mathcal{O}\qty(\gamma^{3/2})\,.
\end{eqnarray}
This beta function features a fixed point at $\gamma=0$ with a critical exponent $\theta_{\gamma} = - \frac{3}{256 \pi^2 }(\beta_+^2+\beta_-^2)$ which is always negative for non-zero $\beta_\pm$. Thus $\gamma$ is marginally irrelevant.
In the case $\beta_+ = \beta_-$ we find that $\gamma=0$ is a fixed line.
 
As we mentioned before, RG flows for $\gamma^{-1}$ have been derived previously, in~\cite{Daum:2010qt}, based on the vielbein and connection as independent degrees of freedom, as well as in~\cite{Benedetti:2011nd} at the perturbative level. Both works found a fixed point at $\gamma=0$, at which $\gamma$ is irrelevant; and a fixed point at $\gamma \rightarrow \pm \infty$, at which $\gamma$ is relevant. From these results, one would conclude that an RG trajectory exists that starts at vanishing parity violation in the deep UV and results in maximal parity violation in the IR.

In our case, $\gamma = 0$ is also always marginally irrelevant. In contrast, whether $\gamma^{-1}=0$ is IR attractive, depends on the sign of the difference between $\beta_+$ and $\beta_-$ as well as the difference between the squares. Because we work in a setting with different degrees of freedom, it is not to be expected that our results match~\cite{Daum:2010qt,Daum:2013fu} exactly. It is, however, interesting that the maximally parity-violating case $\gamma=0$ as well as the absence of parity violating $\gamma \rightarrow \infty$, are both fixed points of all settings.

Because the RG flow for $\gamma^{-1}$ depends on $\beta_{\pm}$, we close this section by considering the RG flow of $\beta_{\pm}$.
In the large-mass limit, the beta functions are given by
\ba
\beta_{\beta_\pm} = \qty(-1 -\frac{3 \beta_\pm^2}{256 \pi^2})\beta_\pm + \mathcal{O}\qty(m_\pm^{-2})
\ea
and in particular depend neither on $\gamma^{-1}$ or $\sigma_{\rho^2}$, 
nor on any other couplings apart from $\beta_\pm$ themselves.

Thus we can integrate these beta functions analytically with the initial conditions $\beta_{\pm}({ k_{\rm UV}})\equiv\beta_{\pm \text{UV}}$ and obtain
\be
\beta_{\pm}^2 (k)=
\frac{ \beta_{\pm  \rm UV}^2}{\left(\frac{k}{k_{\rm UV}}\right)^2\left(1+ \frac{3}{256 \pi^2} \beta_{ \pm \rm UV}^2\right)- \frac{3}{256 \pi^2}\beta_{\pm \rm UV}^2}\,.
\ee 
Under the RG flow to the IR, $\beta_{\pm}$ increase in magnitude and ultimately reach a strong-coupling regime signaled by an IR Landau-pole in the above expression.

We note that, because the masses become large in the IR, one can arrange initial conditions for the couplings and masses such that the IR Landau pole is not reached before the masses become large and shape-mismatching degrees of freedom decouple.

\section{Discussion and outlook}\label{Sec:Discussion}

We have applied FRG techniques to area-metric gravity for the first time. One overarching motivation is to understand whether it is possible to connect area-metric gravity to length-metric gravity in the IR. To do so, we use RG techniques to explore the decoupling of shape-mismatching degrees of freedom. The latter is a necessary condition to recover standard gravity in the IR. 

A second piece of motivation is to connect different approaches to quantum gravity, here spin foams and asymptotic safety. To lay the basis for this connection, we derive the RG flow of three-point vertices that couple length-metric degrees of freedom to shape-mismatching degrees of freedom.

Our results apply to a regime between some fundamental scale  $\Lambda_{\rm UV}$ -- to be understood as the scale at which a more fundamental spin-foam description of spacetime can be written in terms of an effective quantum field theory of length-metric and shape-mismatching degrees of freedom -- and the Planck scale $M_{\rm Planck}$ -- to be understood as the scale at which all quantum-gravitational degrees of freedom decouple. It is a central assumption of our work that such a regime exists.  

In this paper, we focus on our first motivation. We discover indications that shape-mismatching degrees of freedom become massive, with dimensionful masses that are generically $\mathcal{O}(\Lambda_{\rm UV})$, at least in the absence of an extreme fine-tuning of initial conditions. This is a necessary prerequisite for the decoupling of shape-mismatching degrees of freedom and the recovery of GR in the IR.

 We do, however, generically find that the coupling between shape-mismatching and length-metric degrees of freedom also grows towards the IR. This may imply large contributions from off-shell configurations of shape-mismatching degrees of freedom to the effective action for the length metric at scales below the mass scale.

We next  investigate parity symmetry. Shape-mismatching degrees of freedom generically come in a left-handed and right-handed version with different couplings. We discover that the hypersurface at which parity symmetry is restored is IR repulsive at large values of the masses. Thus, unless the theory is fine-tuned so that parity symmetry is (nearly) intact at $\Lambda_{\rm UV}$, parity violation may be large at $M_{\rm Planck}$. This includes parity-violating couplings that couple shape-mismatching degrees of freedom to length-metric degrees of freedom. Thus, despite the decoupling of shape-mismatching degrees of freedom, they may leave an imprint in induced, parity-violating interactions of length-metric degrees of freedom. A more thorough investigation of this point is an obvious future extension of our present work.

Finally, we extract the RG flow of the Immirzi parameter and discover zeros at vanishing Immirzi parameter and vanishing inverse Immirzi parameter. The former is always IR attractive. 
Whether the latter is IR attractive or IR repulsive depends on the coupling of shape-mismatching degrees of freedom with length-metric degrees of freedom.

Our results are subject to several limitations and caveats. First, as is standard in FRG studies, we work in Euclidean signature. In the future, it would be interesting to follow~\cite{Manrique:2011jc,Fehre:2021eob,DAngelo:2023tis,DAngelo:2023wje,Saueressig:2023tfy,Korver:2024sam,Saueressig:2025ypi} and consider Lorentzian RG flows with shape-mismatching degrees of freedom. \\
Second, as is also standard in FRG studies, we truncate the set of interactions. For the future, there are obvious extensions, such as the inclusion of three-vertex interactions of shape-mismatching degrees of freedom, as well the Newton coupling.\\
Third, we are limited in our understanding of the constraints imposed on the space of couplings by diffeomorphism invariance and treat the $h\omega^2$ and $h^2 \omega$ vertices as independent. For the future, understanding geometric invariants based on the area-metric, and writing the action before a perturbative expansion about a flat background, is an important avenue to pursue.

These future extensions will enable us to more robustly understand which properties of gravity in the IR arise from an area-metric setting and whether area-metric gravity can be phenomenologically viable.
As part of such an endeavor, we can in particular connect to the idea of ``effective asymptotic safety". Asymptotic safety is often regarded as a proposal for a UV completion of gravity, but it may -- more conservatively -- be a UV extension that does not hold up to arbitrarily high scales. In such a setting, a more fundamental theory has an effective field theory regime in which the asymptotically safe fixed point occurs as an IR attractive fixed point. This idea has first been put forward in~\cite{Percacci:2010af} and then spelled out with string theory as a possible UV completion in~\cite{deAlwis:2019aud}, see also~\cite{Basile:2021krr}. It generates universality in the sense that multiple initial conditions for the RG flow, corresponding to a variation of free parameters of the UV completion, are mapped to single values (more precisely, very small intervals) at lower scales, see~\cite{Held:2020kze}. Understanding whether there is such a regime of effective asymptotic safety in area-metric gravity is one way of connecting area-metric gravity to standard gravity in the IR. 

In a related question, we can explore whether an asymptotically safe fixed point can even be realized in the presence of shape-mismatching degrees of freedom and not just below their decoupling scale. This will complement the understanding of asymptotic safety in gravity under the impact of matter~\cite{Dona:2013qba, Dona:2014pla, Meibohm:2015twa, Biemans:2017zca, Wetterich:2019zdo, Eichhorn:2022gku} as well as non-standard gravitational degrees of freedom~\cite{Daum:2010qt, Harst:2012ni, Daum:2013fu, Eichhorn:2013xr, Percacci:2013ii, Harst:2015eha}.

Finally, it would be interesting to see whether a coarse graining and renormalization flow in loop quantum gravity and spin foams~\cite{Dittrich:2014ala, Dittrich:2014mxa, Delcamp:2016dqo, Bahr:2016hwc, Asante:2022dnj, Ferrero:2024rvi, Ferrero:2025est} can reproduce features of the functional renormalization flow discussed here for area metrics. As a first step one can stay in the continuum and consider the functional renormalization flow of modified Plebanski actions~\cite{Krasnov:2006du}. Spin foams arise as a quantization of the Plebanski action, the modification accounts for quantization effects, 
 and one can choose a class of modifications, which can be reformulated into area-metric actions~\cite{Borissova:2022clg}. A conjecture in~\cite{Krasnov:2006du} states that the renormalization flow will only produce potential terms involving the shape-mismatching degrees of freedom. With our truncation we stay within this class, but it would be interesting to add terms outside this class in order to test this conjecture.

\begin{acknowledgments}
	
B.~D.~thanks Simone Speziale for discussions. 
Research at Perimeter Institute is supported in part by the Government of Canada through the Department of Innovation, Science and Economic Development Canada and by the Province of Ontario through the Ministry of Colleges and Universities. J.~B.~is supported by an NSERC grant awarded to B.~D.~and a doctoral scholarship by the German Academic Scholarship Foundation. A.~E.~is grateful to Perimeter Institute for hospitality. The work of M.~S.~was in parts supported by a Radboud Excellence fellowship from Radboud University in Nijmegen, Netherlands.
\end{acknowledgments}

\appendix{

\section{Third-order area-metric contractions without derivatives}\label{Appendix:ThirdOrderTerms}

In this appendix we will construct all possible $SO(4)$ invariants at third order in the fields $h$ (the trace of the length metric perturbation), $\hat h$ (the tracefree part of the length metric), $\omega^+$ (the selfdual Weyl component of the area-metric perturbation) and $\omega^-$ (the anti-selfdual Weyl component of the area-metric perturbation). 

To this end we will employ $SO(4)$ representation-theoretic arguments, see \cite{Borissova:2024cpx} for the quadratic case. The fields $h$, $\hat h$, $\omega^+$ and $\omega^-$ are valued in the $(0,0)$, $(1,1)$, $(2,0)$ and $(0,2)$ representations, respectively. To obtain an $SO(4)$ invariant we have to consider tensor products whose decomposition into irreducible components involve the $(0,0)$ representation.

We start with combinations including the trace $h$. In this case the remaining two fields need to couple to $(0,0)$. Thus we obtain all four quadratic invariants, multiplied with $h$:
\ba
h^3, \, 
h(\hat{h}_{\mu\nu})^2 ,h(w^+_{\mu\nu\rho\sigma})^2, \,
h(w^-_{\mu\nu\rho\sigma})^2 \; .
\ea

Next we consider terms with at least two factor of $\hat h$. The tensor product of the corresponding representations is given by 
\ba\label{htimesh}
(1,1)\otimes (1,1)&=& (2,2)\oplus (1,1)\oplus (0,0)  \oplus (2,0) \oplus (0,2) \oplus \nn\\
&&(2,1) \oplus (1,2) \oplus (1,0) \oplus (0,1) \; .
\ea
To construct a third order invariant we need to tensor the irreducible representations appearing on the right hand side with one of the representations appearing in the area metric, so that we can obtain the singlet representation $(0,0)$. We see that there are only four possibilities:

\noindent
Tensoring with $(1,1)$ gives ${\hat h_\mu}^\rho {\hat h_\rho}^\nu {\hat h_\nu}^\mu$.

\noindent
Tensoring with $(0,0)$ we obtain $h(\hat{h}_{\mu\nu})^2$, which was already covered above. 

\noindent
Tensoring with $(2,0)$ we obtain $\hat h^{\mu\nu} \hat h^{\rho\sigma} w^+_{\mu\rho\nu\sigma}$.

\noindent
Tensoring with  $(0,2)$ we obtain  
$\hat h^{\mu\nu} \hat h^{\rho\sigma} w^-_{\mu\rho\nu\sigma}$.

Note that contracting $\hat h^{\mu\nu} \hat h^{\rho\sigma}$ 
with $w^\pm_{\mu\rho\nu\sigma}$ projects out the $(2,0)$ or $(0,2)$ 
part respectively from $\hat h \otimes \hat h$. 
Similarly contracting with $\hat h_{\mu\nu}$ 
projects out the trace-free $(1,1)$ 
part from $\hat h^{\mu\rho} \hat h_\rho^\nu$.

The above coveres all terms which include at least two $\hat h$. Let us move on to third-order terms which are at least quadratic in $w^\pm$. Here we have to consider three tensor products,
\ba\label{wtimesw}
(2,0)\otimes (2,0)&=& (0,0)\oplus(1,0)\oplus (2,0) \oplus  (3,0)\oplus (4,0)\;,\nn\\
(0,2)\otimes (0,2) &=& (0,0)\oplus (0,1)\otimes (0,2) \otimes (0,3)\oplus (0,4)\;,\nn\\
(2,0) \otimes (0,2) &=& (2,2) \; .
\ea
This allows for four third-order terms. We can either multiply $(w^\pm)^2$ with $h$, which gives two terms we already covered. Or we can consider the tensor products $(2,0)\otimes (2,0)\otimes (2,0)$ or $(0,2)\otimes (0,2)\otimes (0,2)$ leading to the two terms
${w^\pm_{\mu\nu}}^{\rho\sigma} {w^\pm_{\rho\sigma}}^{\tau\lambda} {w^\pm_{\tau\lambda}}^{\mu\nu}$.

 To summarize, there are overall 9 invariants of third order in area-metric perturbations without derivatives,
\ba\label{3rdO}
&& h^3, \, 
h(\hat{h}_{\mu\nu})^2 , \,
{\hat h_\mu}^\rho {\hat h_\rho}^\nu {\hat h_\nu}^\mu, \,\nn\\
&& \hat h^{\mu\nu} \hat h^{\rho\sigma} w^+_{\mu\rho\nu\sigma},\, \hat h^{\mu\nu} \hat h^{\rho\sigma} w^-_{\mu\rho\nu\sigma},\, \nn\\
&& h(w^+_{\mu\nu\rho\sigma})^2, \,
h(w^-_{\mu\nu\rho\sigma})^2, \,\nn\\
&& {w^+_{\mu\nu}}^{\rho\sigma} {w^+_{\rho\sigma}}^{\tau\lambda} {w^+_{\tau\lambda}}^{\mu\nu},\, {w^-_{\mu\nu}}^{\rho\sigma} {w^-_{\rho\sigma}}^{\tau\lambda} {w^-_{\tau\lambda}}^{\mu\nu} \, .
\ea

\section{Projectors onto length-metric and non-metric fluctuations}\label{Appendix:Projectors}

The projectors defining the decomposition~\eqref{eq:AParametrization} of the area-metric perturbations $a_{\mu\nu\rho\sigma}$ are given explicitly by
	\ba
	\Pi^\pm_{\mu\nu\rho\sigma,\alpha\beta\gamma\delta} &=& 2 \qty(\mathbb{A}^\pm_{\mu\nu\alpha\beta}\mathbb{A}^\pm_{\rho\sigma\gamma\delta} + \mathbb{A}^\pm_{\mu\nu\gamma\delta}\mathbb{A}^\pm_{\rho\sigma\alpha\beta}) - \frac{4}{3} \mathbb{A}^\pm_{\mu\nu\rho\sigma}\mathbb{A}^\pm_{\alpha\beta\gamma\delta}\,,\\
	\Pi^{\text{L}}_{\mu\nu\rho\sigma,\alpha\beta\gamma\delta} &=& 2\qty(\mathbb{A}^+_{\mu\nu\alpha\beta}\mathbb{A}^-_{\rho\sigma\gamma\delta} + \mathbb{A}^+_{\mu\nu\gamma\delta}\mathbb{A}^-_{\rho\sigma\alpha\beta}) + 2\qty(\mathbb{A}^-_{\mu\nu\alpha\beta}\mathbb{A}^+_{\rho\sigma \gamma\delta} + \mathbb{A}^-_{\mu\nu\gamma\delta} \mathbb{A}^+_{\rho\sigma\alpha\beta}) \nn\\
	&+& \frac{2}{3}\mathbb{A}^{\text{S}}_{\mu\nu\rho\sigma}\mathbb{A}^{\text{S}}_{\alpha\beta\gamma\delta}\,.
	\ea
	where 
	\ba
	\mathbb{A}^\pm_{\mu\nu\rho\sigma} &\equiv & \frac{1}{8}\qty(\delta_{\mu\rho}\delta_{\nu\sigma} - \delta_{\mu\sigma}\delta_{\nu\rho})\pm \frac{1}{8}\epsilon_{\mu\nu\rho\sigma}\,,\\
	\mathbb{A}^{\text{S}}_{\mu\nu\rho\sigma} &\equiv& \mathbb{A}^+_{\mu\nu\rho\sigma} + \mathbb{A}^-_{\mu\nu\rho\sigma} = \frac{1}{4}\qty(\delta_{\mu\rho}\delta_{\nu\sigma} - \delta_{\mu\sigma}\delta_{\nu\rho})\,,\\
	\mathbb{A}^{\text{D}}_{\mu\nu\rho\sigma} &\equiv& \mathbb{A}^{+}_{\mu\nu\rho\sigma} -  \mathbb{A}^{-}_{\mu\nu\rho\sigma} =\frac{1}{4}\epsilon_{\mu\nu\rho\sigma}\,.
	\ea
	Note that by construction
	\ba
	\omega^\pm_{\mu\nu\rho\sigma} &\equiv& \Pi^\pm_{\mu\nu\rho\sigma,\alpha\beta\gamma\delta} a^{\alpha\beta\gamma\delta}\,, \\
	\mathbb{L}^{\lambda\tau}_{\mu\nu\rho\sigma} h_{\lambda\tau} &\equiv & \Pi^{\text{L}}_{\mu\nu\rho\sigma,\alpha\beta\gamma\delta} a^{\alpha\beta\gamma\delta}
	\ea
	and moreover
	\be
	h_{\lambda\tau} = \mathbb{K}_{\lambda\tau,\alpha\beta\gamma\delta}a^{\alpha\beta\gamma\delta} 
	\ee
	where
	\be
	 \mathbb{K}_{\lambda\tau,\alpha\beta\gamma\delta} \equiv \frac{1}{8} \mathbb{I}_{\lambda\tau \lambda'\tau'}  \mathbb{L}^{\lambda'\tau'}_{\alpha\beta\gamma\delta} -\frac{1}{2\cdot 3} \delta_{\lambda\tau}\mathbb{A}^{\text{S}}_{\alpha\beta\gamma\delta}\,,\,\,\, \mathbb{I}_{\lambda\tau \lambda'\tau'} = \frac{1}{2}\qty(\delta_{\lambda \lambda'}\delta_{\tau\tau'} + \delta_{\lambda \tau'}\delta_{\tau\lambda'})\,.
	\ee
The Projectors $\Pi^{\mathrm{TT}}$ and $\Pi^0$, which project onto the spin-2 and spin-0 components of metric fluctuations respectively are defined by
\begin{align}
 {\Pi^{\mathrm{TT}}}\indices{_{\mu\nu}^{\rho\sigma}}=\,&\delta\indices{_{(\mu}^{\rho}}\,\delta\indices{_{\nu)}^{\sigma}}-\frac{1}{3}\delta_{\mu\nu}\,\delta^{\rho\sigma}-\frac{2}{p^2}\delta\indices{_{(\mu}^{(\rho}}\,p\indices{_{\nu)}}\,p\indices{^{\sigma)}}\\
 &+\frac{1}{3}\frac{1}{p^2}\left(\delta_{\mu\nu}\, p^{\rho}\, p^{\sigma}+p_{\mu} \, p_{\nu} \, \delta^{\rho\sigma}\right)+ \frac{2}{3}\frac{1}{p^4} p_{\mu} \,p_{\nu}\, p^{\rho} \,p^{\sigma}\,,\notag\\
{\Pi^{\mathrm{0}}}\indices{_{\mu\nu}^{\rho\sigma}}=&\,\frac{(-3+\beta_h)^2}{12(3+\beta_h^2)} \delta_{\mu\nu} \delta^{\rho\sigma} +\frac{4\beta_h^2}{(9+3\beta_h^2)}\frac{1}{p^4}\,p_{\mu}\,p_{\nu}\,p^{\rho}\,p^{\sigma}\\
&+\frac{(-3+\beta_h)\beta_h}{3(3+\beta_h^2)}\frac{1}{p^2}(p_{\mu}\,p_{\nu}\,\delta^{\rho\sigma}+\delta_{\mu\nu}\,p^{\rho}\,p^{\sigma})\,,
\end{align}
where the explicit gauge-dependence of $\Pi^0$ ensures that the propagating scalar mode is projected on irrespective of the choice of $\beta_h$, see \cite{Knorr:2021slg, Knorr:2021niv}. For $\beta_h\to0$, $\Pi^0$ reduces to the projector on the trace mode.
\bibliographystyle{jhep}
\bibliography{references}

\end{document}